\documentclass[aps,preprint,superscriptaddress,showpacs,12pt]{revtex4-1}
\usepackage{dcolumn}
\usepackage{comment}
\usepackage[dvips]{graphicx}
\begin{document}
\title{\textbf{The damping and diffusion of atoms moving in the background electromagnetic environment}}
\author{Li Ge}
\affiliation{School of Science, Hangzhou Dianzi University, Hangzhou, 310018, China}
\email{geli0922@hdu.edu.cn}

\begin{abstract}
The interaction between an atom and the quantized electromagnetic field depends on the position of the atom. Then the atom experiences
a force which is the minus gradient of this interaction. Through the Heisenberg equations of motion and the Born-Markov
approximation, the mean and correlation of the force are obtained, showing that the center-of-mass motion of the atom is damped and diffused.
This approach can be easily generalized to multi-level atoms, where the damping force and diffusion coefficients are just the weighted average of the contributions from all pairs
of energy levels that have nonvanishing dipole elements. It is shown that these results are invariant under Galilean
transformation, and in principle can be used to determine the velocity of the lab relative to the background radiation.
\end{abstract}


\maketitle
\section{introduction}
The interaction between light and atom has been studied in quite detail since the revolution of quantum mechanics. The understanding
of this interaction is crucial to the development of laser and optical manipulation techniques~\cite{ashkin1,ashkin2}, which stimulated remarkable advances in various
fields of physics~\cite{wieman,ketterle,bloch,vahala,girvin1,favero,genes1,kiesel,schwab,roels,raizen}.
It has been realized that the electromagnetic background (or the 'vacuum' at zero temperature) is responsible for
many effects such as spontaneous emission, Casimir forces and Lamb shift etc. In quantum optics, the electromagnetic background is often regarded as an
environment that interacts with the atom and leads to dephasing and dissipation. These effects, which can be characterized by several decay rates,
are important for optical cooling since finally the energy of the atom is dissipated to the environment.

In any optical manipulation scheme, the force exerted on the atom is mainly due to the laser-atom interaction. Besides, the background field also induces
an damping force, though quite small in most situations. Qualitatively this is like an object moves in the air and it experiences a wind.
It's easy to see that this force vanishes at zero temperature since the vacuum is Lorentz invariant and we can always choose a frame where the atom is at rest.
At finite temperature, however, the thermal background is not invariant under such coordinate transformation and when the atom moves, it experiences a wind
of photons that produces a damping force. Due to momentum conservation, the momentum of the atom changes once it absorbs or emits a photon.
Then an equation governing the evolution of the momentum distribution of the atoms can be deduced, showing that the atomic motion is damped
and diffused  ~\cite{cohen}.

The derivation is, however, a little tedious and there remain two problems: one is to generalize the results to multilevel atoms, and the other is, since the lab frame may be in relative motion with the background(for example, earth moves relative to the cosmic background), how does the damping force changes under transformation of frame?
In the theory of laser-cooling, there is often another approach, that is, the damping force and diffusion coefficients are determined by the the mean and correlation of the force operator
~\cite{cohen1,cohen2}.
In this paper we generalize this approach to take into account the interaction of the atom and electromagnetic environment (as dynamical variables).
Through this approach, we first reproduce the result of two level atom and then generalize it to multilevel atoms.
Finally we consider the case that the lab is in relative motion with the background and show that the damping force and diffusion
coefficient are invariant under Galilean transformation.

\subsection{Outline of the model}
The whole system consists of a two-level atom and the electromagnetic environment, which is described by the Hamiltonian:
$H=H_A+H_c+H_F+V$, where $H_A=\frac{\hbar\omega_0}{2}\sigma_z$
is the internal atomic Hamiltonian, $H_c=p^2/2M$ is the center-of-mass Hamiltonian, and
\begin{equation}
H_F=\sum_{k\lambda} \hbar\omega_k a_{k\lambda}^\dagger a_{k\lambda}
\end{equation}
is the Hamiltonian for the free electromagnetic field, with $a_{k\lambda}$ ($a_{k\lambda}^{\dag}$) the annihilation (creation) operator for
the mode of momentum $k$ and polarization $\lambda$.
Under the dipole and rotating wave approximation, the light-atom interaction is:
\begin{equation} \label{interact}
V =-\mathbf{\hat{d}}\cdot\mathbf{E}(\mathbf{R})\approx-i\sum_{k\lambda} \sqrt{\frac{\hbar\omega_k}{2\epsilon_0\Omega}}(\mathbf{d}\cdot\mathbf{\varepsilon}_{k\lambda})
(a_{k\lambda}\sigma_+e^{i\mathbf{k}\cdot \mathbf{R}}-a_{k\lambda}^\dagger \sigma_{-}e^{-i\mathbf{k}\cdot \mathbf{R}})
\end{equation}
where $\mathbf{\hat{d}}=\mathbf{d}\sigma_x$  is the dipole operator, \textbf{R} is the position of the atom.
This interaction leads to the absorption and emission of photons as well as transition between atomic internal levels, and since $V$ depends on
$\mathbf{R}$, it also influences the center of mass motion of the atom. An earlier treatment of this problem is to consider the evolution of the momentum distributions  $\pi_a(\mathbf{p})$ and $\pi_b(\mathbf{p})$, which represent the occupation number in the internal level $a$ or $b$ with total momentum $\mathbf{p}$.  Each time the atom absorbs (emits) a photon with momentum $\hbar \mathbf{k}$, its own momentum
changes with the same amount, so the time derivative of $\pi_a(\mathbf{p})$ contains a term proportional to $\pi_b(p+\hbar k)$, and similar is for $\pi_b(\mathbf{p})$.
In the case that $\hbar k$ is much smaller than the width $\Delta p$ of momentum distributions, one has $\pi_a(\mathbf{p}\pm\hbar \mathbf{k})\approx\pi_a(\mathbf{p})\pm\hbar \mathbf{k}\cdot \nabla \pi_a(\mathbf{p})$ and a differential equation governing the evolution
of  $\pi_a(\mathbf{p})$ and $\pi_b(\mathbf{p})$ can be obtained.
 Commonly, the characteristic evolution time of the external variable is much longer than that of the internal variables, then
$\pi_a(\mathbf{p})$ and $\pi_b(\mathbf{p})$ adapt themselves quasi- instantaneously to the much slower variations of the total distribution $\pi(\mathbf{p})=\pi_a(\mathbf{p})+\pi_a(\mathbf{p})$, giving $\pi_a(\mathbf{p})=\frac{1}{1+e^{\beta\hbar\omega_0}}\pi(\mathbf{p}),\pi_b(\mathbf{p})=\frac{e^{-\beta\hbar\omega_0}}{1+e^{\beta\hbar\omega_0}}\pi(\mathbf{p})$.
Finally one gets the evolution equation of $\pi(p)$, which is a Fokker-Planck type equation.
It can be shown that $d \overline{\mathbf{p}}/dt=-\gamma\overline{\mathbf{p}}$ and $d\overline{p^2}/dt=-2\gamma\overline{p^2}+2D$,
where $\overline{\mathbf{p}}, \overline{p^2}$ are ensemble averages over the distribution $\pi(\mathbf{p})$ ,and $\gamma, D$ are the damping and diffusion coefficients, respectively~\cite{cohen}.

In this paper we exploit another approach which is more straightforward compared to the previous work. From the Heisenberg equation of motion:
\begin{equation}
\frac{d\mathbf{p}}{dt}=\frac{1}{i\hbar}[\mathbf{p},H]=-\nabla V\equiv\langle\mathbf{F}\rangle+\mathbf{\xi}
\end{equation}
  where the force operator  $\mathbf{F}=-\nabla V$ is divided into two parts: $\langle\mathbf{F}\rangle$ is the expectation value of $\mathbf{F}$  and $\mathbf{\xi}$ is the fluctuation. This is a Langevin equation describing the Brownian motion of the atom, and the theory of Brownian motion naturally tells that $\gamma, D$ are determined by the  mean and correlation
  of $\mathbf{F}$~\cite{breuer}. When the atom is at rest, the mean force is obviously 0, then to the first order of $\mathbf{p}$ one has 
  $\langle\mathbf{F}\rangle=-\gamma \mathbf{p}=-\gamma M\mathbf{v}$. On the other hand, the diffusion coefficient is: $D_{\alpha \beta}=\frac{1}{2} \int_{-\infty}^\infty \langle \xi_\alpha(t)\xi_\beta(t')\rangle dt'$, where $\alpha,\beta=x,y,z$ denotes the spatial component.
  Similar approach has also been used in the theory of laser cooling~\cite{cohen1,cohen2}, where the laser is treated as classical electromagnetic field.

From (\ref{interact}) we get:
\begin{eqnarray}
\mathbf{F}&=&-\sum_{k\lambda} \mathbf{k}g_{k\lambda}(a_{k\lambda}\sigma_+e^{i\mathbf{k}\cdot \mathbf{R}}+a_{k\lambda}^\dagger \sigma_{-}e^{-i\mathbf{k}\cdot \mathbf{R}}) \nonumber \\
&=& -\frac{1}{2}\sum_{k\lambda}\mathbf{k}g_{k\lambda}[(a_{k\lambda}\sigma_{+}+\sigma_{+}a_{k\lambda})e^{i\mathbf{k}\cdot \mathbf{R}}+(a_{k\lambda}^\dagger\sigma_{-}+\sigma_{-}a_{k\lambda}^\dag) e^{-i\mathbf{k}\cdot \mathbf{R}}]
\end{eqnarray}
where $g_{k\lambda}=(\mathbf{d}\cdot\mathbf{\varepsilon}_{k\lambda})\sqrt{\frac{\hbar\omega_k}{2\epsilon_0\Omega}}$ and in the second equality
the symmetric ordering between the atom and field operators is chosen~\cite{cohen3,cohen4}.
To calculate the expectation and correlation of any operator $\hat{O}(t)$ for the atomic internal variables or field variables,
we need first know  the quantum state of the 'atom$+$field' system. For an atom  weakly coupled to
the environment with huge degrees of freedom, the state of the environment remains almost unaffected and the Born-Markov approximation can be made, under which the density matrix of
the whole system is $\rho(t)\simeq\rho_A(t)\otimes \frac{1}{Z_F}e^{-\beta H_F}$~\cite{breuer,cohen}. Starting from the Liouville equation $i\hbar\partial_t \rho=[H,\rho]$, the
master equation of $\rho_A(t)$ is obtained by tracing over the environment variables and keeping terms up to the second order of $V$.  The steady state
of the atom is $\frac{1}{Z_A}e^{-\beta H_A}$, which is in thermal equilibrium with the environment. Commonly, the characteristic evolution time of the external variable is much longer than that of the internal variables~\cite{cohen}, then adiabatic approximation can be taken which assumes the internal state is in the steady state, i.e. the expectation values $\langle\hat{O}\rangle$ for any atomic internal variables or field variables is taken over the density matrix  $\frac{1}{Z} e^{-\beta H_0}$, where $H_0=H_A+H_F$.

We need also solve the time evolution of the operators.
It is convenient to introduce $\widetilde{a}_{k\lambda}=a_{k\lambda}e^{i\mathbf{k}\cdot \mathbf{R}}$, then
 \begin{equation} \label{force}
\mathbf{F}=-\frac{1}{2\hbar}\sum_{k\lambda} g_{k\lambda}\mathbf{k}(\widetilde{a}_{k\lambda}\sigma_{+}+\sigma_{+}\widetilde{a}_{k\lambda}+\widetilde{a}_{k\lambda}^\dagger\sigma_{-}+\sigma_{-}\widetilde{a}_{k\lambda}^\dag)
\end{equation}
and the Heisenberg equations of motion are:
\begin{eqnarray}
i\hbar\frac{d\widetilde{a}_{k\lambda}}{dt}&=&(\omega_k-\mathbf{k}\cdot \frac{d\mathbf{R}}{dt})\widetilde{a}_{k\lambda}+i\frac{g_{k\lambda}}{\hbar}\sigma_- \nonumber\\
i\frac{d\sigma_-}{dt}&=&\frac{\omega_0}{2}\sigma_- +i\frac{g_{k\lambda}}{\hbar}\sigma_z\widetilde{a}_{k\lambda} \nonumber \\
i\frac{d\sigma_+}{dt}&=&-\frac{\omega_0}{2}\sigma_+ +i\frac{g_{k\lambda}}{\hbar}\sigma_z\widetilde{a}_{k\lambda}^\dag
\end{eqnarray}
with the solutions:
\begin{eqnarray} \label{evolve}
\widetilde{a}_{k\lambda}&=&\widetilde{a}_{k\lambda}(0)e^{-i[\omega_kt-\mathbf{k}\cdot(\mathbf{R}(t)-\mathbf{R}(0))]}+\frac{g_{k\lambda}}{\hbar}\int_0^t
e^{-i[\omega_\mathbf{k}(t-s)-\mathbf{k}\cdot (\mathbf{R}(t)-\mathbf{R}(s)]}\sigma_{-}(s) ds \nonumber \\
\sigma_{-}(t)&=&\sigma_{-}(0)e^{-i\omega_0t}+\frac{g_{k\lambda}}{\hbar}\int_0^t
e^{-i[\omega_0(t-s)]}\sigma_{z}(s)\widetilde{a}_{k\lambda}(s) ds  \nonumber \\
\sigma_{+}(t)&=&\sigma_{+}(0)e^{i\omega_0t}+\frac{g_{k\lambda}}{\hbar}\int_0^t
e^{i[\omega_0(t-s)]}\sigma_{z}(s)\widetilde{a}_{k\lambda}^\dag(s) ds
\end{eqnarray}
These equations contain two parts: the free evolution terms and the source terms proprotional to the interaction strength
$g_{k\lambda}$. Apparently the products of two free terms contribute nothing to $\langle\mathbf{F}\rangle$, and the leading contributions  come from terms like $g_{k\lambda}^2 \sigma_-(s)\sigma_+(t)$.
Then in the spirit of Born-Markov approximation we can make the approximation $\hat{O}(t)=e^{iHt}\hat{O}(0)e^{-iHt}\approx e^{iH_0t}\hat{O}(0)e^{-iH_0t}$, since the effect of $V$ is at least of the order $g_{k\lambda}^3$.

\subsection{The mean and correlation of $\mathbf{F}$}

From above we have
\begin{eqnarray}
\langle \mathbf{F}\rangle &=& -\frac{1}{2\hbar}\sum_{k\lambda} g_{k\lambda}^2\mathbf{k}\int_0^t
e^{-i[\omega_k(t-s)-\mathbf{k}\cdot (\mathbf{R}(t)-\mathbf{R}(s)]}(\langle\sigma_{-}(s)\sigma_+(t)\rangle+\langle\sigma_+(t)\sigma_{-}(s)\rangle  ds+c.c. \nonumber \\
& -& \frac{1}{2\hbar}\sum_{k\lambda} g_{k\lambda}^2\mathbf{k}\int_0^t
e^{-i[(\omega_k-\omega_0)(t-s)-\mathbf{k}\cdot (\mathbf{R}(t)-\mathbf{R}(s)]}
\langle\sigma_{z}\rangle(\langle\widetilde{a}_{k\lambda}(0)\widetilde{a}_{k\lambda}^\dag(0)\rangle+\langle\widetilde{a}_{k\lambda}^\dag(0)\widetilde{a}_{k\lambda}(0)\rangle) ds+c.c. \nonumber \\
&=&-\frac{1}{2\hbar} \sum_{k\lambda} g_{k\lambda}^2\mathbf{k}\int_0^t
e^{-i[(\omega_k-\omega_0)(t-s)-\mathbf{k}\cdot (\mathbf{R}(t)-\mathbf{R}(s)]}
[1+\langle\sigma_{z}\rangle(2n_k+1)] ds+c.c.
\end{eqnarray}
where $\langle\sigma_{-}(s)\sigma_+(t)\rangle+\langle\sigma_+(t)\sigma_{-}(s)\rangle= e^{i\omega_0(t-s)} $ has been used. The integrand becomes negligible as long as
$t-s\gg\tau_c$, where $\tau_c$ is the correlation time of the environment, so for slowly moving atom we can approximate
$\mathbf{R}(t)-\mathbf{R}(s)$ as $\mathbf{v} (t-s)$ and let the lower limit of the integral go to minus infinity~\cite{breuer}.
 Finally:
\begin{eqnarray}
\langle \mathbf{F}\rangle
= -\frac{1}{2\hbar}\sum_{k\lambda} g_{k\lambda}^2\mathbf{k}\int_{-\infty}^t
e^{-i[(\omega_k-\omega_0-\mathbf{k}\cdot\mathbf{v})(t-s)}
[1+\langle\sigma_{z}\rangle(2n_k+1)] ds+c.c.\nonumber \\
=-\frac{\pi}{\hbar} \sum_{k\lambda}  g_{k\lambda}^2\mathbf{k} \delta(\omega_k-\omega_0-\mathbf{k}\cdot\mathbf{v})[1+\langle\sigma_{z}\rangle(2n_k+1)]
\end{eqnarray}
where $n_k=1/(e^{\beta\hbar\omega_k}-1)$ is the Bose distribution function and $\langle\sigma_{z}\rangle=-1/(2n(\omega_0)+1)$. In reality what we consider is an ensemble of atoms,
 where the directions of $\mathbf{d}$ is distributed uniformly. Then by taking the ensemble average:
 \begin{equation}\label{ave}
   \sum_\lambda \overline{g_{k\lambda}^2}=\frac{\hbar\omega_k}{2\epsilon_0\Omega}\sum_{\lambda,i,j}\overline{d_id_j}\mathbf{\varepsilon}_{k\lambda}^i\mathbf{\varepsilon}_{k\lambda}^j
   =\frac{\hbar\omega_k}{2\epsilon_0\Omega}\sum_{i,j}\frac{d^2\delta_{ij}}{3}(1-\frac{k_ik_j}{k^2})=\frac{\hbar\omega_k}{\epsilon_0\Omega}\frac{d^2}{3}
 \end{equation}
This procedure is equivalent to the isotropic radiation pattern taken in~\cite{cohen}.
Next by transforming the summation to integral: $\Sigma_k\rightarrow\Omega/(2\pi)^3 \int d^3k $ and expanding the $\delta$ function to the first order
of $\mathbf{k}\cdot\mathbf{v}$, we have:
\begin{eqnarray}
\langle \mathbf{F}\rangle
= -\frac{\pi d^2}{3\epsilon_0}\int \frac{d^3k}{(2\pi)^3} \omega_k\mathbf{k}
 [\delta(\omega_k-\omega_0)+\frac{d\delta(\omega_k-\omega_0)}{d\omega_k}(\mathbf{k}\cdot\mathbf{v})][1+\langle\sigma_{z}\rangle(2n_k+1)]
\end{eqnarray}
The first term is the force exerted on a rest atom, which is obviously 0, so
\begin{eqnarray} \label{tforce}
\langle \mathbf{F}\rangle
&=& \frac{\pi d^2}{3\epsilon_0}\int \frac{d^3k}{(2\pi)^3}\delta(\omega_k-\omega_0)\frac{d}{d\omega_k}\big(\omega_k\mathbf{k}(\mathbf{k}\cdot\mathbf{v})[1+\langle\sigma_{z}\rangle(2n_k+1)]\big) \nonumber\\
&=& -\frac{\pi c d^2k_0^3N(\omega_0)}{9\epsilon_0}\frac{dn(\omega_0)/d\omega_0}{2n(\omega_0)+1}\mathbf{v}
\end{eqnarray}
where $k_0=\omega_0/c$, $N(\omega_0)$ is the density of states of the photon. Using the formular of the spontaneous emission rate: $\Gamma=\frac{\pi\omega_0d^2N(\omega_0)}{3\hbar \epsilon_0}$, we have $\langle \mathbf{F}\rangle=-\gamma \mathbf{p}=-\gamma M\mathbf{v}$, with
\begin{equation}
\gamma=\frac{\hbar k_0^2\Gamma}{3M}\frac{dn(\omega_0)/d\omega_0}{2n(\omega_0)+1}
\end{equation}
which is  identical with that in~\cite{cohen}.

Next let's turn to the force correlation function. Up to the order of $g_{k\lambda}^2$, it is:
\begin{equation}
  \langle F_\alpha(t)F_\beta(t')\rangle=\Sigma_{k\lambda}\overline{g_{k\lambda}^2}k_\alpha k_\beta\big(\langle a_{k\lambda}(t)a^\dag_{k\lambda}(t')\rangle\langle\sigma_+(t)\sigma_-(t')\rangle
  +\langle a^\dag_{k\lambda}(t)a_{k\lambda}(t')\rangle\langle\sigma_-(t)\sigma_+(t')\rangle \big)
\end{equation}
 The momentum diffusion coefficient is:
\begin{equation}
  D_{\alpha \beta}=\frac{1}{2} \int_{-\infty}^\infty \langle \xi_\alpha(t)\xi_\beta(t')\rangle dt'
  =\frac{1}{2} \int_{-\infty}^\infty \big[\langle F_\alpha(t)F_\beta(t')\rangle- \langle F_\alpha(t)\rangle\langle F_\beta(t')\rangle\big] dt'
\end{equation}
The second term is of the order $g_{k\lambda}^4$ and can be neglected. Using $\langle a_{k\lambda}(t)a^\dag_{k\lambda}(t')\rangle=(1+n_k)e^{-i\omega_k(t-t')}$,
$\langle\sigma_+(t)\sigma_-(t')\rangle=\frac{n(\omega_0)}{2n(\omega_0)+1}e^{i\omega_0(t-t')}$,
 $\langle\sigma_-(t)\sigma_+(t')\rangle=\frac{n(\omega_0)+1}{2n(\omega_0)+1}e^{-i\omega_0(t-t')}$, it can be shown that:
\begin{eqnarray}
D_{\alpha \beta}&\approx& \frac{1}{2} \int_{-\infty}^\infty \langle F_\alpha(t)F_\beta(t')\rangle dt'
=\frac{\hbar d^2}{18\epsilon_0}\delta_{\alpha\beta} \int_{-\infty}^\infty \frac {d^3k}{(2\pi)^3}
\int_{-\infty}^\infty d\tau\omega_kk^2 \frac{n(\omega_0)+n_k+2n_kn(\omega_0)}{2n(\omega_0)+1}e^{i(\omega_0-\omega_k)\tau} \nonumber \\
&=&\frac{2\pi\hbar d^2}{9\epsilon_0}\frac{n(\omega_0)+n^2(\omega_0)}{2n(\omega_0)+1} \delta_{\alpha\beta} \int_{-\infty}^\infty \frac {d^3k}{(2\pi)^3} \omega_kk^2\delta(\omega_0-\omega_k) =\frac{\pi\hbar cd^2k_0^3}{9\epsilon_0}\frac{n(\omega_0)+n^2(\omega_0)}{2n(\omega_0)+1}N(\omega_0)\delta_{\alpha\beta} \nonumber \\
\end{eqnarray}
In term of the spotaneous emission rate $\Gamma$, it is
\begin{equation}
  D_{\alpha \beta}=\frac{\hbar^2 k_0^2\Gamma}{3}\frac{n(\omega_0)+n^2(\omega_0)}{2n(\omega_0)+1}
\end{equation}
which is also identical with that in~\cite{cohen}.

\subsection{Generalization to multi-level atoms}
The above calculation can be easily generalized to the case of multi-level atoms, where the atomic Hamiltonian is:
\begin{equation}
H_A=\sum_{i}\hbar\omega_i|i\rangle\langle i|
\end{equation}
and the atom-light interaction:
\begin{equation}
V=-i\sum_{k\lambda,ij} \sqrt{\frac{\hbar\omega_k}{2\epsilon_0\Omega}}(\mathbf{d}_{ij}\cdot\mathbf{\varepsilon}_{k\lambda})
(a_{k\lambda}\sigma_+^{ij}e^{i\mathbf{k}\cdot \mathbf{R}}-a_{k\lambda}^\dagger \sigma_{-}^{ij}e^{-i\mathbf{k}\cdot \mathbf{R}})
\end{equation}
with $\sigma_+^{ij}=|i\rangle\langle j|$ and $\sigma_-^{ij}=|j\rangle\langle i|$. The summation are taken over all pairs of states $(i,j)$ with nonzero
dipole element $\mathbf{d}_{ij}=\langle i|\mathbf{\hat{d}}|j\rangle$.

Similarly we introduce $\widetilde{a}_{k\lambda}=a_{k\lambda}e^{i\mathbf{k}\cdot \mathbf{R}}$, then following the derivation of two-level system, we have:
\begin{eqnarray} \label{f2}
\widetilde{a}_{k\lambda}&=&\widetilde{a}_{k\lambda}(0)e^{-i[\omega_kt-\mathbf{k}\cdot(\mathbf{R}(t)-\mathbf{R}(0))]}+\sum_{ij}\frac{g_{k\lambda,ij}}{\hbar}\int_0^t
e^{-i[\omega_\mathbf{k}(t-s)-\mathbf{k}\cdot (\mathbf{R}(t)-\mathbf{R}(s)]}\sigma_{-}^{ij}(s) ds \nonumber \\
\sigma_{-}^{ij}(t)&=&\sigma_{-}^{ij}(0)e^{-i\omega_{ij}t}+\frac{g_{k\lambda,ij}}{\hbar}\int_0^t
e^{-i[\omega_{ij}(t-s)]}\sigma_{z}^{ij}(s)\widetilde{a}_{k\lambda}(s) ds  \nonumber \\
\sigma_{+}^{ij}(t)&=&\sigma_{+}^{ij}(0)e^{i\omega_{ij}t}+\frac{g_{k\lambda,ij}}{\hbar}\int_0^t
e^{i[\omega_{ij}(t-s)]}\sigma_{z}^{ij}(s)\widetilde{a}_{k\lambda}^\dag(s) ds
\end{eqnarray}
where $\omega_{ij}=\omega_{i}-\omega_{j}$, $g_{k\lambda,ij}=(\mathbf{d}_{ij}\cdot\mathbf{\varepsilon}_{k\lambda})\sqrt{\frac{\hbar\omega_k}{2\epsilon_0\Omega}}$

And the force operator is:
\begin{equation}
\mathbf{F}=-\frac{1}{2}\sum_{k\lambda,ij}\mathbf{k} g_{k\lambda,ij}(\widetilde{a}_{k\lambda}\sigma_{+}^{ij}+\sigma_{+}^{ij}\widetilde{a}_{k\lambda}
+\widetilde{a}_{k\lambda}^\dagger\sigma_{-}^{ij}+\sigma_{-}^{ij}\widetilde{a}_{k\lambda}^\dag)
\end{equation}

To calculate the mean of $\mathbf{F}$, we encounter terms like $g_{k\lambda,ij}^2\langle \sigma_{+}^{ij}(t)\sigma_{-}^{i'j'}(s)\rangle$.
Remember that the average are taken over the density matrix  $\frac{1}{Z} e^{-\beta H_0}$, the above terms can be nonzero only for $i'=i, j'=j$, then
\begin{eqnarray} \label{mforce}
\langle \mathbf{F}\rangle
&=& -\frac{1}{2\hbar} \sum_{k\lambda} \sum_{ij}\overline{g_{k\lambda,ij}^2}\mathbf{k}\int_0^t
e^{-i[(\omega_k-\omega_{ij}-\mathbf{k}\cdot\mathbf{v})(t-s)}
[\rho_{ii}+\rho_{jj}+\langle\sigma_{z}^{ij}\rangle(2n_k+1)] ds+c.c. \nonumber\\
&=& \sum_{ij}\frac{\pi d^2}{3\epsilon_0}\int \frac{d^3k}{(2\pi)^3}\delta(\omega_k-\omega_{ij})(\rho_{ii}+\rho_{jj})\frac{d}{d\omega_k}
\big(\omega_k\mathbf{k}(\mathbf{k}\cdot\mathbf{v})[1+\langle\sigma^{ij}_{z}\rangle(2n_k+1)]\big) \nonumber\\
&=& -\sum_{ij}\frac{\pi c d^2k_{ij}^3N(\omega_{ij})}{9\epsilon_0}\frac{dn(\omega_{ij})/d\omega_{ij}}{2n(\omega_{ij})+1}(\rho_{ii}+\rho_{jj})\mathbf{v}
\end{eqnarray}
where $k_{ij}=\omega_{ij}/c$ and $\rho_{ii}=e^{-\beta\hbar\omega_i}/\sum_i(e^{-\beta\hbar\omega_i})$ is the occupation probability of the $i$th level.
Comparing  (\ref{mforce}) with (\ref{tforce}), it's easy to see that the damping force for a multilevel atom is just the weighted average of the contributions from all pairs
of energy levels that have nonvanishing dipole elements. This is also true for the diffusion coefficient. Proceeding the calculations we get

\begin{equation}
D_{\alpha \beta}=\sum_{ij}\frac{\pi\hbar^2 c d^2k_{ij}^3N(\omega_{ij})}{9\epsilon_0}\frac{n(\omega_{ij})
[n(\omega_{ij})+1]}{2n(\omega_{ij})+1}(\rho_{ii}+\rho_{jj})\delta_{\alpha \beta}
\end{equation}

\subsection{Invariance of $\gamma$ and $D$ under Galilean transformation}

The above results are what measured from the background frame. In principle, the lab can be in
relative motion with the background, then a problem naturally arises: what are these coefficients measured in the lab? Usually, the force is invariant
under transformation of coordinate frame in nonrelativistic mechanics, and as will be seen later, it is also true in this case.
 Suppose the lab moves with a velocity $\mathbf{u}$ relative to the background,
then the Galilean transformation between the two frame is $t'=t, \mathbf{R}'=\mathbf{R}-\mathbf{u}t$. The electric field in the lab frame is also transformed as
$\mathbf{E}'=\mathbf{E}-\mathbf{u}\times \mathbf{B}$, where $\mathbf{B}$ is the magnetic field in the background frame. As like $\mathbf{E}$, $\mathbf{B}$ has the plane wave
expansion:
\begin{equation}\label{magnetic}
  \mathbf{B}=i\sum_{k\lambda} \sqrt{\frac{\hbar}{2\epsilon_0\omega_k\Omega}}(\mathbf{k}\times\mathbf{\varepsilon}_{k\lambda})
(a_{k\lambda}e^{i\mathbf{k}\cdot \mathbf{R}}-a_{k\lambda}^\dagger e^{-i\mathbf{k}\cdot \mathbf{R}})
\end{equation}
The ratio  between the expansion coefficients in $\mathbf{B}$  and  $\mathbf{E}$ is: $k/\omega_k=1/c$, so in the nonrealtivistic limit the $\mathbf{u}\times \mathbf{B}$
term can be neglected and $\mathbf{E}'\approx \mathbf{E}$. Also under the Galilean transformation, $\nabla'=\nabla$, then the force operator in the lab frame is
 $\mathbf{F}'=\nabla'(\mathbf{\hat{d}}\cdot\mathbf{E'})= \nabla(\mathbf{\hat{d}}\cdot\mathbf{E})= \mathbf{F}$, where we have used the fact that $\hat{d}$
 is invariant under Galilean transformation.
It is also known that the exact density matrix of the electromagnetic environment in the lab frame is $\rho_0=\frac{1}{Z_0} e^{-\beta u^\mu P_\mu}$, where $u^\mu$
is the 4-velocity of the lab, $P_\mu=\sum_{k\lambda}\hbar k^\mu a_{k\lambda}^\dagger a_{k\lambda}$ is the 4-momentum of the field~\cite{hakim}.
 This formula is explicitly invariant under Lorentz transformation,
 so it is of course invariant under Galilean transformation if the nonrelativistic limit is taken. Combining all these together, we reach the conclusion that
$\gamma$ and $D$ are invariant under Galilean transformation. 
Then in the lab frame, the atom experiences a damping force
 \begin{equation}\label{lab}
   \langle \mathbf{F'}\rangle=\langle \mathbf{F}\rangle=-\gamma M \mathbf{v}=-\gamma M(\mathbf{u}+\mathbf{v}')
 \end{equation}
where $\mathbf{v}'$ is the velocity of the atom relative to the lab. Under the action of this force, the final mean velocity of the atoms
will be $-\mathbf{u}$ in the lab frame. To keep the atoms move with the lab, an extra force equals $\gamma M\mathbf{u}$ must be exerted to the atoms,
and in principle, the measurement of this force can be used to determine the velocity of the lab relative to the background radiation.

\subsection{Conclusion}
In  conclusion, we derive the damping and diffusion coefficients for atoms moving in the background electromagnetic field.
 The center of mass motion of the atom is described by a Langevin equation, and the theory of Brownian motion immediately gives the general forms
 of the damping and diffusion coefficients. This approach is widely used in the theory of laser cooling, so our work shows that the optical force produced by a laser or by the background field can be calculated in a unified way. The generalization to multilevel atoms is straightforward, and it's easy to show these coefficients are invariant under Galilean transformation.

 \section{acknowledgements}
  This work is supported by Natural Science Foundation of Zhejiang Province LQ18A040003.
%


\begin{thebibliography}{99}
\bibitem{ashkin1}A. Ashkin, Phys. Rev. Lett. {\bf 24}, 156 (1970).
\bibitem{ashkin2} A. Ashkin, {\em Optical Trapping and Manipulation of Neutral Particles Using Lasers} (World Scientific, Singapore, 2006).
\bibitem {wieman} M. H. Anderson, J. R. Ensher, M. R. Matthews, C. E. Wieman, and E. A. Cornell, Science {\bf 269}, 198 (1995).
\bibitem{ketterle} K. B. Davis, M.-O. Mewes, M. R. Andrews, N. J. van Druten, D. S. Durfee,
D. M. Kurn, and W. Ketterle, Phys. Rev. Lett. {\bf 75}, 3969 (1995).
\bibitem{bloch} I. Bloch, J. Dalibard, and W. Zwerger, Rev. Mod. Phys. {\bf 80}, 885 (2008).
\bibitem{vahala} T. Kippenberg and K. Vahala, Science {\bf 321}, 1172 (2008).
\bibitem{girvin1} F. Marquardt and S. Girvin, Physics {\bf 2}, 40 (2009).
\bibitem{favero} I. Favero and K. Karrai, Nat. Photon. {\bf 3}, 201 (2009).
\bibitem{genes1} C. Genes, A.Mari, D. Vitali, and P. Tombesi, Adv. At.Mol. Opt. Phys. {\bf 57}, 33 (2009).
\bibitem{kiesel} M. Aspelmeyer, S. Groblacher, K. Hammerer, and N. Kiesel, J. Opt. Soc. Am. B {\bf 27}, A189 (2010).
\bibitem{schwab} M. Aspelmeyer and K. Schwab, New J. Phys. {\bf 10}, 095001 (2008).
\bibitem{roels} D. van Thourhout and J. Roels, Nat. Photon. {\bf 4}, 211 (2010).
\bibitem{raizen} T. Li, S. Kheifets, and M. G. Raizen, Nat. Phys. {\bf 7}, 527 (2011).
\bibitem{cohen} C. Cohen-Tannodji, J. Dupont-Roc, G.Grynberg, {\em Atom-photon interactions: basic processes and applications} (Wiley-VCH Verlag GmbH and Co. KGaA, 2004).
\bibitem{cohen1} J. Dalibard and C. Cohen-Tannoudji, J. Opt. Soc. Am. B {\bf 6}, 2023 (1989).
\bibitem{cohen2} J. Dalibard, S. Reynaud and C. Cohen-Tannoudji, J. Phys. B: At. Mol. Phys. {\bf 17}, 4577 (1984).
\bibitem{cohen3} J. Dalibard, J. Dupont-Roc, C. Cohen-Tannodji, J. Phys. (France) {\bf 43}, 1617 (1982).
\bibitem{cohen4} J. Dalibard, J. Dupont-Roc, C. Cohen-Tannodji, J. Phys. (France) {\bf 45}, 637 (1984).
\bibitem{breuer} H. P. Breuer, F. Petruccione, {\em The Theory of Open Quantum Systems} (Oxford University Press, 2002).
\bibitem{hakim} R. Hakim, {\em Introduction to Relativistic Statistical Mechanics, Classical and Quantum} (World Scientific Publishing Co. Pte. Ltd, 2011).
\end{thebibliography}
\end{document}